\documentclass[10pt,conference]{IEEEtran}
\usepackage{amsfonts}
\usepackage{graphicx}
\usepackage{amsmath}
\usepackage{amssymb}
\usepackage{verbatim}
\newtheorem{thm}{Theorem}

\begin{document}
\title{Noisy DPC and Application to a Cognitive Channel}
\author{
\authorblockN{Yong Peng and Dinesh Rajan, \emph{Senior Member, IEEE}}
}
\maketitle \footnotetext[1]{The authors are with the Department of
Electrical Engineering, Southern Methodist University, Dallas, TX,
USA. Email: \{ypeng,rajand\}@lyle.smu.edu. This work has been
supported in part by the National Science Foundation through grant CCF
0546519.}

\begin{abstract}
In this paper, we first consider a channel that is contaminated by
two independent Gaussian noises $S\sim \mathcal{N}(0,Q)$ and
${Z_0}\sim\mathcal{N}(0,N_0)$. The capacity of this channel is
computed when independent noisy versions of ${S}$ are known to the
transmitter and/or receiver. It is shown that the channel capacity
is greater then the capacity when ${S}$ is completely unknown, but
is less then the capacity when ${S}$ is perfectly known at the
transmitter or receiver. For example, if there is one noisy version
of ${S}$ known at the transmitter only, the capacity is
$\frac{1}{2}\log(1+\frac{P}{Q(N_1/(Q+N_1))+N_0})$, where $P$ is the
input power constraint and $N_1$ is the power of the noise
corrupting $S$. We then consider a Gaussian cognitive interference
channel~(IC) and propose a causal noisy dirty paper coding (DPC)
strategy. We compute the achievable region using this noisy DPC
strategy and quantify the regions when it achieves the upper bound
on the rate.
\end{abstract}
\section{INTRODUCTION}

Consider a channel in which the received signal, $\mathbf{Y}$ is
corrupted by two independent additive white Gaussian noise (AWGN)
sequences,~$\mathbf{S}\sim \mathcal{N}(0,Q\mathbf{I}_n)$ and
$\mathbf{Z}_0\sim\mathcal{N}(0,N_0\mathbf{I}_n)$, where
$\mathbf{I}_n$ is the identity matrix of size~$n$. The received
signal is of the form,
\begin{equation}
\mathbf{Y} = \mathbf{X} + \mathbf{S} + \mathbf{Z}_0,
\label{eq:chan_model}
\end{equation}
where $\mathbf{X}$ is the transmitted sequence for $n$ uses of the
channel. Let the transmitter and receiver each has knowledge of
independent noisy observations of~$\mathbf{S}$.  We quantify the
benefit of this additional knowledge by computing the capacity of the
channel in~(\ref{eq:chan_model}) and presenting the coding scheme that
achieves capacity. Our result indicates that the capacity is of the
form $C(\frac{P}{\mu Q + N_0})$, where $C(x)=0.5 \log(1+x)$ and $0\le
\mu \le 1$ is the residual fraction~(explicitly characterized in
Sec.~\ref{sec:capacity}) of the interference power, $Q$, that can not
be canceled with the noisy observations at the transmitter and
receiver.

We then consider the network in Fig.~\ref{fig:DPC_cog} in which the
primary transmitter~(node~$A$) is sending information to its intended
receiver~(node~$B$). There is also a secondary transmitter~(node~$C$)
who wishes to communicate with its receiver~(node~$D$) on the same
frequency as the primary nodes. We focus on the case when nodes~$C$
and~$D$ are relatively closer to node~$A$ than node~$B$.  Such a
scenario might occur for instance when node~$A$ is a cellular base
station and nodes~$C$ and~$D$ are two nearby nodes, while node~$B$ is
at the cell-edge.

Let node~$A$ communicate with its receiver node~$B$ at rate~$R$
using transmit power~$P_A$. Let the transmit power of node~$C$
equal~$P_C$. Since we assumed that node~$B$ is much farther away
from the other nodes, we do not explicitly consider the interference
that $P_C$ causes at node~$B$. A simple lower bound, $R_{CD-lb}$ on
the rate that nodes~$C$ and ~$D$ can communicate is
\begin{equation}
R_{CD-lb}= C({|h_{CD}|^2P_C}/{(N_D+|h_{AD}|^2 P_A)}),
\end{equation}
which is achieved by treating the signal from node~$A$ as noise at
node~$D$. Similarly, a simple upper bound on this rate is obtained~(if
either nodes~$C$ or~$D$ has perfect, noncausal knowledge of
node~$A$'s signal) as
\begin{equation}
R_{CD-ub}= C({|h_{CD}|^2P_C}/{N_D}).
\label{eq:rate_ub_cog}
\end{equation}
We propose a new causal transmission scheme based on the noisy DPC
strategy derived in Sec.~\ref{sec:noisyDPC}. This new scheme
achieves the upper bound~(\ref{eq:rate_ub_cog}) in some scenarios,
which are quantified.

\section{Noisy Dirty Paper Coding \label{sec:noisyDPC}}

\subsection{System Model\label{sec:model}}
\begin{figure}[htp]
\begin{center}
\includegraphics[width=3.0in]{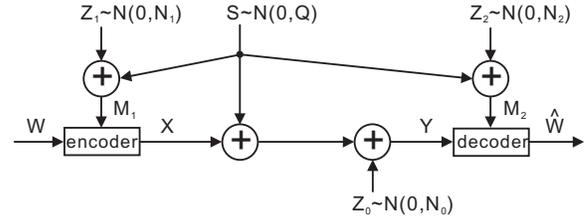}
\caption{A channel with noise observed at both encoder and decoder.}
\label{fig:DPC_1}
\end{center}
\end{figure}
The channel model is depicted in Fig. \ref{fig:DPC_1}. The transmitter
sends an index, $W\in \{1,2,\ldots,K\}$, to the receiver in $n$ uses
of the channel at rate $R = \frac{1}{n}\log_2K$ bits per
transmission. The output of the channel in~(\ref{eq:chan_model}) is
contaminated by two independent AWGN sequences, $\mathbf{S}\sim
\mathcal{N}(0,Q\mathbf{I}_n)$ and
$\mathbf{Z}_0\sim\mathcal{N}(0,N_0\mathbf{I}_n)$. Side information
$\mathbf{M}_1=\mathbf{S}+\mathbf{Z}_1$, which is noisy observations of
the interference is available at the transmitter. Similarly, noisy
side information $\mathbf{M}_2=\mathbf{S}+\mathbf{Z}_2$, is available
at the receiver. The noise vectors are distributed as $\mathbf{Z}_1
\sim \mathcal{N}(0,N_1 \mathbf{I}_n)$ and $\mathbf{Z}_2 \sim
\mathcal{N}(0,N_2 \mathbf{I}_n)$.

Based on index $W$ and $\mathbf{M_1}$, the encoder transmits one
codeword, $\mathbf{X}$, from a $(2^{nR},n)$ code book, which satisfies
average power constraint, $\frac{1}{n}\|\mathbf{X}\|^2 \leq P$. Let
$\hat{W}$ be the estimate of~$W$ at the receiver; an error occurs if
$\hat{W}\neq W$.
\subsection{Related Work}
One special case of~(\ref{eq:chan_model}) is when a noisy version of
$\mathbf{S}$ is known only to the transmitter; our result in this case
is a generalization of Costa's celebrated result~\cite{Costa}.
In~\cite{Costa}, it is shown that the achievable rate when the noise
$\mathbf{S}$ is perfectly known at the transmitter is equivalent to
the rate when $\mathbf{S}$ is known at the receiver, and this rate
does not depend on the variance of $\mathbf{S}$. A new coding strategy
to achieve this capacity was also introduced in~\cite{Costa} and is
popularly referred to as dirty paper coding~(DPC). We generalize
Costa's result to the case of noisy interference knowledge. We show
that the capacity with knowledge of a noisy version of $\mathbf{S}$ at
the transmitter is equal to the capacity with knowledge of a
statistically equivalent noisy version of $\mathbf{S}$ at the
receiver. However, unlike~\cite{Costa} where the capacity does not
depend on the variance of~$\mathbf{S}$, in the general noisy side
information case, the capacity decreases as the variance
of~$\mathbf{S}$ increases.

In~\cite{Costa}, Costa adopted the random coding argument given
by~\cite{Gelfand,Gamal}. Based on the channel capacity
$C=\max_{p(u,x|s)}\{I(U;Y)-I(U,S)\}$ given in \cite{Gelfand,Gamal},
Costa constructed the auxiliary variable~${U}$ as a linear
combination of ${X}\sim \mathcal{N}(0,P)$ and ${S}\sim
\mathcal{N}(0,Q)$ and showed that this simple construction of~${U}$
achieves capacity.

Following Costa's work, several extensions of DPC have been studied,
\emph{e.g.}, colored Gaussian noise~\cite{Yu}, arbitrary
distributions of~$\mathbf{S}$~\cite{Cohen} and deterministic
sequences~\cite{Erez}. The case when $\mathbf{S}$ is perfectly known
to the encoder and a noisy version is known to the decoder is
considered in~\cite{Mazzotti}, mainly focusing on discrete
memoryless channels. The only result in~\cite{Mazzotti} for Gaussian
channel reveals no additional gain due to the presence of the noisy
estimate at the decoder, since perfect knowledge is available at the
encoder and DPC can be used. In contrast, in this paper we study the
case when only noisy knowledge of~$\mathbf{S}$ is available at both
transmitter and receiver.
\subsection{Channel Capacity}\label{sec:capacity}
\begin{thm}
Consider a channel of the form~(\ref{eq:chan_model}) with an average
transmit power constraint~$P$. Let independent noisy observations
$\mathbf{M}_1=\mathbf{S}+\mathbf{Z}_1$
and~$\mathbf{M}_2=\mathbf{S}+\mathbf{Z}_2$ of the
interference~$\mathbf{S}$ be available, respectively, at the
transmitter and receiver. The noise vectors have the following
distributions: $\mathbf{Z}_i \sim \mathcal{N}(0,N_i \mathbf{I}_n)$,
$i=0,1,2$ and $\mathbf{S} \sim \mathcal{N}(0,Q \mathbf{I}_n)$. The
capacity of this channel equals $C\left(\frac{P}{\mu
Q+N_0}\right)$, where $0\le
\mu=\frac{1}{1+\frac{Q}{N_1}+\frac{Q}{N_2}} \le 1$.
\label{thm:DPC}
\end{thm}
\emph{Remark:} Clearly $\mu= 0$ when either $N_1=0$ or $N_2=0$ and the
capacity is $C(P/N_0)$, which is consistent
with~\cite{Costa}\footnote{Costa's result is a special case with
$N_1=0$ and $N_2=\infty$.}. Further, $\mu=1$ when $N_1\rightarrow
\infty$ and $N_2\rightarrow \infty$, and the capacity is
$C(P/(Q+N_0))$, which is the capacity of a Gaussian channel with
noise~$Q+N_0$.  Thus, one can interpret~$\mu$ as the residual
fractional power of the interference that cannot be canceled by the
noisy observations at the transmitter and receiver.

\emph{Proof:} We first compute an outer bound on the capacity of
this channel. It is clear that the channel capacity can not exceed
$\max_{p(x|m_1,m_2)}I(X;Y|M_1,M_2)$, which is the capacity when both
$M_1$ and $M_2$ are known at the transmitter and receiver. Thus, a
capacity bound of the channel can be calculated as
\begin{align}
&~I(X;Y|M_1,M_2) = I(X;Y,M_1,M_2) - I(X;M_1,M_2) \nonumber \\
\leq&~ I(X;Y,M_1,M_2)\label{eq:cap_ub_step}  \\
=&~H(X)+H(Y,M_1,M_2)-H(X,Y,M_1,M_2)\nonumber
\end{align}
\begin{align}
=&~{\frac{1}{2}\log(2\pi e)^4P\left|\begin{array}{ccc}
P+Q+N_0 & Q  & Q\\
Q & Q+N_1 & Q\\
Q & Q & Q+N_2
\end{array} \right|}\nonumber\\
&~-{\frac{1}{2}\log(2\pi e)^4\left|\begin{array}{cccc}
P & P & 0 & 0\\
P & P+Q+N_0 & Q  & Q\\
0 & Q & Q+N_1 & Q\\
0 & Q & Q & Q+N_2\\
\end{array} \right|} \nonumber
\end{align}
\begin{align}
=C\left({P}/{(\mu
Q+N_0)}\right).\quad\quad\quad\quad\quad\quad\quad\quad\quad\quad\quad~~
\label{eq:capacity}
\end{align}
where $\mu=\frac{1}{1+\frac{Q}{N_1}+\frac{Q}{N_2}}$. Note that the
inequality in~(\ref{eq:cap_ub_step}) is actually a strict equality
since $I(X;M_1,M_2)=0$.

\subsection{Achievability of Capacity}\label{sec:Achieve}
We now prove that~(\ref{eq:capacity}) is achievable. The codebook
generation and encoding method we use follows the principles
in~\cite{Gelfand, Gamal}. The construction of auxiliary variable is
similar to~\cite{Costa}.

\emph{Random codebook generation:}

1) Generate $2^{nI(U;Y,M_2)}$ i.i.d. length-$n$ codewords
$\mathbf{U}$, whose elements are drawn i.i.d. according to
$U\sim\mathcal{N}(0,P+\alpha^2(Q+N_1))$, where $\alpha$ is a
coefficient to be optimized.

2) Randomly place the $2^{nI(U;Y,M_2)}$ codewords $\mathbf{U}$ into
$2^{nR}$ cells in such a way that each of the cells has the same
number of codewords. The codewords and their assignments to the
$2^{nR}$ cells are revealed to both the transmitter and the
receiver.

\emph{Encoding:}

1) Given an index $W$ and an observation,
$\mathbf{M_1}=\mathbf{M_1}(i)$, of the Gaussian noise sequence,
$\mathbf{S}$, the encoder searches among all the codewords
$\mathbf{U}$ in the $W^{th}$ cell to find a codeword that is jointly
typical with $\mathbf{M_1}(i)$. It is easy to show using the joint
asymptotic equipartition property (AEP)~\cite{Cover} that if the
number of codewords in each cell is at least $2^{nI(U,M_1)}$, the
probability of finding such a codeword $\mathbf{U}=\mathbf{U}(i)$
exponentially approaches $1$ as $n\rightarrow \infty$.

2) Once a jointly typical pair $(\mathbf{U}(i),\mathbf{M_1}(i))$ is
found, the encoder calculates the codeword to be transmitted as
$\mathbf{X}(i)=\mathbf{U}(i)-\alpha \mathbf{M_1}(i)$. With high
probability, $\mathbf{X}(i)$ will be a typical sequence which
satisfies $\frac{1}{n}\|\mathbf{X}(i)\|^2\leq P$.

\emph{Decoding:}

1) Given $\mathbf{X}(i)$ is transmitted, the received signal is
$\mathbf{Y}(i)=\mathbf{X}(i)+\mathbf{S}+\mathbf{Z_0}$. The decoder
searches among all $2^{nI(U;Y,M_2)}$ codewords $\mathbf{U}$ for a
sequence that is jointly typical with $\mathbf{Y}(i)$. By joint AEP,
the decoder will find $\mathbf{U}(i)$ as the only jointly typical
codeword with probability approaching 1.

2) Based on the knowledge of the codeword assignment to the cells,
the decoder estimates $\hat{W}$ as the index of the cell that
$\mathbf{U}(i)$ belongs to.

\emph{Proof of achievability:}

 Let $U=X+\alpha M_1=X+\alpha
(S+Z_1)$, $Y=X+S+Z_0$ and $M_2=S+Z_2$, where $X\sim
\mathcal{N}(0,P)$, $S\sim \mathcal{N}(0,Q)$ and $Z_i\sim
\mathcal{N}(0,N_i),~i=0,1,2$ are independent Gaussian random
variables. To ensure that with high probability, in each of the
$2^{nR}$ cells, at least one jointly typical pair of $\mathbf{U}$
and $\mathbf{M_1}$ can be found. The rate,~$R$, which is a function
of $\alpha$, must satisfy
\begin{equation}
R(\alpha)\leq I(U;Y,M_2)-I(U;M_1).\label{eq:rate}
\end{equation}
The two mutual informations in~(\ref{eq:rate}) can be calculated as
\begin{align}
&~I(U;Y,M_2) = ~H(U)+H(Y,M_2)-H(U,Y,M_2)\nonumber\\
=&~\frac{1}{2}\log\left({\left[P+\alpha^2(Q+N_1)\right]\left|\begin{array}{cc}
P+Q+N_0 & Q \\
Q & Q+N_2
\end{array}\right|}\right)\label{eq:I1}\\
-&\frac{1}{2}\log\left({\left|\begin{array}{ccc}
P+\alpha^2(Q+N_1) & P+\alpha Q  & \alpha Q\\
P+\alpha Q & P+Q+N_0 & Q\\
\alpha Q & Q & Q+N_2
\end{array}\right|}\right)\nonumber
\end{align}
\begin{equation}
\mbox{and } I(U;M_1)=\frac{1}{2}\log\left(\frac{P+\alpha^2(Q+N_1)}{P}\right).\label{eq:I2}
\end{equation}
Substituting~(\ref{eq:I1}) and~(\ref{eq:I2}) into~(\ref{eq:rate}), we find
\begin{align}
&R(\alpha) \le \frac{1}{2} \log P[(Q+P+N_0)(Q+N_2)-Q^2 ]
\nonumber\\
&- \frac{1}{2}\log
\left\{\alpha^2[Q(P+N_0)(N_1+N_2)+(Q+P+N_0)N_1N_2]\right.\nonumber\\
&\left.-2\alpha QPN_2+P(QN_0+QN_2+N_0N_2) \right\}. \label{eq:rate1}
\end{align}
After simple algebraic manipulations, the optimal coefficient,
$\alpha^*$, that maximizes the right hand side of (\ref{eq:rate1})
is found to be
\begin{equation}
\alpha^*=
\frac{QP{{N_2}}}{Q(P+N_0)({N_1}+{N_2})+(Q+P+N_0){N_1N_2}}.\label{eq:alpha}
\end{equation}
Substituting for $\alpha^*$ in~(\ref{eq:rate1}), the maximal rate
equals
\begin{equation}
R(\alpha^*)=C\left({P}/{(\mu
Q+N_0)}\right)\label{eq:achieve}
\end{equation}
with $\frac{1}{\mu}={1+\frac{Q}{N_1}+\frac{Q}{N_2}}$, which equals the
upper bound~(\ref{eq:capacity}).

\subsection{Special cases}\label{sec:generalization}
{\underline{Noisy estimate at transmitter/receiver only:}}
When the observation of $\mathbf{S}$ is only available at the
transmitter or receiver, the channel is equivalent to our original
model when $N_2\rightarrow \infty$ and $N_1\rightarrow \infty$,
respectively. Their capacity are, respectively
\begin{align}
I(X;Y|M_1)&=C({P}/{(Q[N_1/{(Q+N_1)}]+N_0)})\label{eq:capacity_1}\\
I(X;Y,M_2)&=C({P}/{(Q[N_2/{(Q+N_2)}]+N_0)}),
\label{eq:C_rx}
\end{align}

Note that when $N_1=0$, the channel model further reduces to Costa's
DPC channel model~\cite{Costa}. This paper extends that result to
the case of noisy interference. Indeed, by setting $N_1=N_2$ in
(\ref{eq:C_rx}) and (\ref{eq:capacity_1}), we can see that the
capacity with noisy interference known to transmitter only equals
the capacity with a statistically similar noisy interference known
to receiver only.

From~(\ref{eq:capacity_1}), one may intuitively interpret the
effect of knowledge of~$M_1$ at the transmitter. Indeed, a fraction
$\frac{Q}{Q+N_1}$ of the interfering power can be canceled using the
proposed coding scheme. The remaining $\frac{N_1}{Q+N_1}$ fraction
of the interfering power, $Q$, is treated as `residual' noise. Thus,
unlike Costa's result~\cite{Costa}, the capacity in this case
depends on the power~$Q$ of the interfering source: For a
fixed~$N_1$, as~$Q \rightarrow \infty$, the capacity decreases and
approaches $C\left({P}/{(N_1+N_0)}\right)$.

{\underline{Multiple Independent Observations:}}
Let there be $n_1$ independent observations
$\mathbf{M}_{1},\mathbf{M}_2,\ldots,$$\mathbf{M}_{n_1}$ of
$\mathbf{S}$ at the transmitter and $n_2$ independent observations
$\mathbf{M}_{n_1+1},$$\mathbf{M}_{n_1+2},$$\ldots,\mathbf{M}_{n_1+n_2}$
at the receiver. It can be easily shown that the capacity in this case
is given by $C\left({P}/{(\hat{\mu}Q+N_0)}\right)$, where
$\hat{\mu}=\frac{1}{1+\frac{Q}{N_1}+\frac{Q}{N_2}+\cdots
+\frac{Q}{N_{n_1+n_2}}}$ and $N_1,N_2,\ldots,N_{n_1+n_2}$ are the
variances of the Gaussian noise variables, corresponding to the
$n_1+n_2$ observations.  The proof involves calculating maximum
likelihood estimates~(MLE) of the interference at both the transmit
and receive nodes and using these estimates in
Theorem~\ref{thm:DPC}. To avoid repetitive derivations, the proof is
omitted.

It is easy to see that the capacity expression is symmetric in the
noise variances at the transmitter and receiver. In other words,
having all the $n_1+n_2$ observations at the transmitter would result
in the same capacity. Thus, the observations of $\mathbf{S}$ made at
the transmitter and the receiver are equivalent in achievable rate, as
long as the corrupting Gaussian noises have the same statistics.

In this section, we assumed non-causal knowledge of the interference
at the transmitter and receiver nodes. In the next section, we propose
a simple and practical transmission scheme that uses causal knowledge
of the interference to increase the achievable rate.

\section{Applying DPC to a Cognitive Channel \label{sec:cog_cap}}
\begin{figure}[htp]
\begin{center}
\includegraphics[width=2.5in]{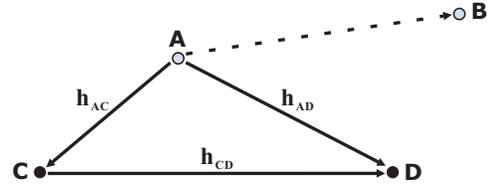}
\caption{Cognitive interference channel model.} \label{fig:DPC_cog}
\end{center}
\end{figure}
\begin{thm}
Consider the network as shown in Fig.~\ref{fig:DPC_cog}. Nodes~$C$
can communicate with node~$D$ at rate given by~(\ref{eq:rate_cogic})
\begin{figure*}
\begin{equation}
R_{CD} = \begin{cases}
C(\frac{|h_{CD}|^2P_C}{N_D}) & \mbox{ if } |h_{AD}|^2 \ge \frac{P_C |h_{CD}|^2+ N_D}{P_A} (e^{2R}-1) \\
C(\frac{|h_{CD}|^2P_C}{(N_D+P_A|h_{AD}|^2)}) & \mbox{ if } |h_{AD}|^2 \le \frac{N_D}{P_A}(e^{2R}-1) \mbox{and }
|h_{AC}|^2 \le \frac{N_C}{P_A}(e^{2R}-1) \\
C (\frac{|h_{CD}|^2P_C }{\mu_{r} |h_{AD}|^2P_A + N_D}) & \mbox{ if } |h_{AC}|^2 \le \frac{N_C}{P_A} (e^{2R}-1) \mbox{ and }
 \frac{P_C |h_{CD}|^2+ N_D}{P_A} (e^{2R}-1) \ge |h_{AD}|^2 \ge \frac{N_D}{P_A}(e^{2R}-1) \\
(1-\frac{m}{n}) C (\frac{|h_{CD}|^2P_C(n/n-m) }{\mu_{t} |h_{AD}|^2P_A + N_D}) & \mbox{ if } |h_{AC}|^2 \ge \frac{N_C}{P_A}(e^{2R}-1)
 \mbox{and} |h_{AD}|^2 \le \frac{N_D}{P_A}(e^{2R}-1) \\
(1-\frac{m}{n}) C (\frac{|h_{CD}|^2P_C(n/n-m) }{\mu_{tr} |h_{AD}|^2P_A + N_D}) & \mbox{ if } |h_{AC}|^2 \ge \frac{N_C}{P_A}(e^{2R}-1)
 \mbox{and} \frac{P_C |h_{CD}|^2+ N_D}{P_A} (e^{2R}-1) \ge |h_{AD}|^2 \le \frac{N_D}{P_A}(e^{2R}-1)
\end{cases}
\label{eq:rate_cogic}
\end{equation}
\end{figure*}
where $\mu_r = \frac{1}{1+0.5 e^{nE_D(R)}}$, $\mu_t = \frac{1}{1+0.5 e^{mE_C(R)}}$ and $\mu_{tr} = \frac{1}{1+0.5 e^{mE_C(R)}+ 0.5 e^{nE_D(R)}}$.
\end{thm}

{\em Proof:} Consider the various cases as follows:

1. Let $|h_{AD}|^2 \ge \frac{P_C |h_{CD}|^2+ N_D}{P_A}
(e^{2R}-1)$. Now, consider the multiple access channel from
nodes~$A$,~$C$ to node~$D$. Clearly, node~$D$ can decode the signal
transmitted by node~$A$ by treating the signal from node~$C$ as
noise. Hence, it can easily subtract this signal from the received
signal and node~$C$ can achieve its rate upper
bound~$C(P_C|h_{CD}|^2/N_C)$.

2. Consider the case $|h_{AD}|^2 \le \frac{N_D}{P_A}(e^{2R}-1)$ and
$|h_{AC}|^2 \le \frac{N_C}{P_A}(e^{2R}-1)$. Now, neither node~$C$ nor
node~$D$ can perfectly decode the signal from node~$A$. Thus, an
achievable rate of $C(\frac{|h_{CD}|^2P_C}{(N_D+P_A|h_{AD}|^2)})$ for
node~$C$ is obtained simply by treating the signal from node~$A$ as
noise at node~$D$.

3. Now, consider the case $|h_{AC}|^2 \ge \frac{N_C}{P_A}(e^{2R}-1)$ and
$\frac{P_C |h_{CD}|^2+ N_D}{P_A} (e^{2R}-1) \ge |h_{AD}|^2 \ge
\frac{N_D}{P_A}(e^{2R}-1)$ In the following we construct a simple
practical scheme in which nodes~$C$ and~$D$ obtain causal, noisy
estimates of the signal being sent from node~$A$. Using these
estimates and Theorem~\ref{thm:DPC}, the nodes cancel out a part of
the interference to achieve a higher transmission rate as follows.

Let us assume that node~$A$ uses a code book of size~$(2^{nR},n)$
where each element is \emph{i.i.d.} Gaussian distributed. The transmit
signal is denoted as $X_{A}(i), i = 1, 2, \ldots n$.  Nodes~$C$
and~$D$ listen to the signal transmitted by node~$A$ for $m$~symbols
in each block of~$n$ symbols. Based on the received signal, nodes~$C$
and~$D$ decodes the code word transmitted by node~$A$. Let~$P_{e,C}$
and $P_{e,D}$ denote, respectively, the probability of decoding error
at nodes~$C$ and~$D$: These error probabilities depend on the channel
gains as well as~$m$. In the remaining~$n-m$ symbols, nodes~$C$
and~$D$ use their estimate of $X_{A}(i), i=m+1, \ldots n$ to increase
their transmission rate. Using Theorem~\ref{thm:DPC}, the achievable
rate is given by
\begin{equation}
r=\frac{1}{2}\left(1-\frac{m}{n}\right) \log \left(1 + \frac{|h_{CD}|^2P_C(n/n-m) }{\mu_{tr} |h_{AD}|^2P_A + N_D}\right),
\label{eq:rate_nDPC_cog}
\end{equation}
where
\begin{equation}
\frac{1}{\mu_{tr}} = 1 + \frac{|h_{AD}|^2P_A}{N_1} + \frac{|h_{AD}|^2P_A}{N_2}
\label{eq:mu_tr}
\end{equation}
The transmit power at node~$C$ is increased over the $n-m$ symbols
that it transmits to meet average power constraint~$P_C$. The variance
of error in the estimate of~$X_A$ at nodes~$C$ and~$D$ is given
respectively by $N_1$ and~$N_2$. Because of the \emph{i.i.d} Gaussian
code book being used, $N_1 = 2P_{e,C} P_A |h_{AD}|^2$ and $N_2 = 2
P_{e,D} P_A |h_{AD}|^2$. The value of $P_{e,C}$ and $P_{e,D}$ can be
obtained using the theory of error exponent.  Specifically, using the
random coding bound, we obtain,
\begin{equation}
P_{e,C} \le \exp(-m E_{C}(R)) \mbox{ and } P_{e,D} \le \exp(-n
E_{D}(R))
\end{equation}
where $E_C(R)$ and $E_D(R)$ represent the random coding
exponent. $E_C(R)$ is derived in~\cite{Gallager1968} and shown
in~(\ref{eq:error_exp1}) for easy reference~($E_D(R)$ is similarly
defined).
\begin{figure*}
\begin{equation}
E_C(R) = \begin{cases}
0 & \text{if}~R > C\left(\frac{|h_{AC}|^2P_C}{N_C}\right)\\
\frac{A_1}{4\beta}
\left[(\beta+1)-(\beta-1)\sqrt{1+\frac{4\beta}{A_1(\beta-1)}}~\right]+\frac{1}{2}\log
\left(\beta-\frac{A_1(\beta-1)}{2}\left[\sqrt{1+\frac{4\beta}{A_1(\beta-1)}}-1\right]\right)
& \text{if}~\delta \le R \le C \left( \frac{|h_{AC}|^2P_C}{N_C}\right)\\
1-\gamma+\frac{A_1}{2} +\frac{1}{2} \log \left(\gamma-\frac{A_1}{2}\right)+
\frac{1}{2} \log (\gamma) - R&\text{if}~ R < \delta
\end{cases}
\label{eq:error_exp1}
\end{equation}
\end{figure*}
In~(\ref{eq:error_exp1}), $A_1=\frac{|h_{AC}|^2P_{A}}{N_C}$, $\beta =
exp(2R)$, $\gamma= 0.5(1+\frac{A_1}{2}+\sqrt{1+\frac{A_1^2}{4}})$,
$\delta=0.5
\log(0.5+\frac{A_1}{4}+0.5\sqrt{1+\frac{A_1^2}{4}})$. Substituting
for~$N_1$ and~$N_2$ into~(\ref{eq:mu_tr}), one can obtain the rate
given in~(\ref{eq:rate_cogic}).

Note that there is no constraint that node~$C$ must use codes of
length~$m-n$ since node~$A$ uses codes of length~$n$. Node~$C$ can
code over multiple codewords of~$A$ to achieve its desired probability
of error.

The selection of~$m$ critically affects the achievable rates. On the
one hand, increasing~$m$ results in lesser fraction of time available
for actual data communications between nodes~$C$ and~$D$ and thus
decreasing rate. On the other hand, increasing~$m$ results in improved
decoding of node~$A$'s signal at nodes~$C$ and~$D$ consequently
reducing~$P_{e,C}$ and $P_{e,D}$ and increasing the achievable
rate. The optimal value of~$m$ can be obtained by equating the
derivative of~(\ref{eq:rate_nDPC_cog}) to 0. Due to the analytical
intractability, we resort to simple numerical optimization to find the
optimal value of~$m$. For a given~$n$, we evaluate the rate~$r_{CD}$
for all values of $m =1,2,\ldots n$ and then simply pick the largest
value.  We are currently trying to derive analytical expressions for
the optimum value of~$m$.

4. Let $|h_{AC}|^2 \le \frac{N_C}{P_A} (e^{2R}-1)$ and
$\frac{P_C |h_{CD}|^2+ N_D}{P_A} (e^{2R}-1) \ge |h_{AD}|^2 \ge
\frac{N_D}{P_A}(e^{2R}-1)$. In this case, the transmitter node~$C$
cannot decode node~$A$'s signal. However, node~$D$ uses all~$n$
received symbols to first decode node~$A$'s signal (with certain
error probability) and then cancel its effect from the received
signal. Subsequently, node~$D$ will decode node~$C$'s signal and the
achievable rate is obtained from Theorem~\ref{thm:DPC}.

5. Finally, let $|h_{AC}|^2 \ge \frac{N_C}{P_A}(e^{2R}-1)$ and
$|h_{AD}|^2 \le \frac{N_D}{P_A}(e^{2R}-1)$. In this case, node~$D$
cannot decode node~$A$'s signal. However, node~$C$ uses the
first~$m$ received symbols to first decode node~$A$'s signal (with
certain error probability) and then employ a noisy DPC transmission
strategy. Subsequently, the achievable rate is obtained from
Theorem~\ref{thm:DPC}. $\hfill \square$

\subsection{Numerical Results}

In our numerical results we fix the values for the parameters as:
$P_A=10$, $P_C=2, N_C=N_D=1$. For simplicity we fix $|h_{CD}|=1$ and
vary $h_{AC}$ and $h_{AD}$.

\begin{figure}[htp]
\begin{center}
\includegraphics[width=3in]{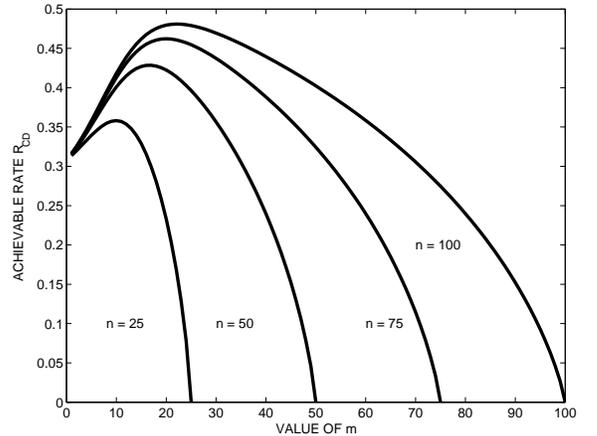}
\caption{Variation of achievable rate with~$m$ for different values
of~$n$.} \label{fig:rate_varn_m}
\end{center}
\end{figure}
Fig.~\ref{fig:rate_varn_m} shows the variation of the achievable rate
with~$m$ for different values of~$n$. As $n$ increases the fractional
penalty on the rate for larger~$m$ is offset by the gains due to
better decoding. Thus, the optimum value of~$m$ increases. However, it
turns out that the optimum ratio~$m/n$ decreases as~$n$ increases.  We
are currently trying to analytically compute the limit to which the
optimum~$m$ converges as $n \rightarrow \infty$.

\begin{figure}[htp]
\begin{center}
\includegraphics[width=3in]{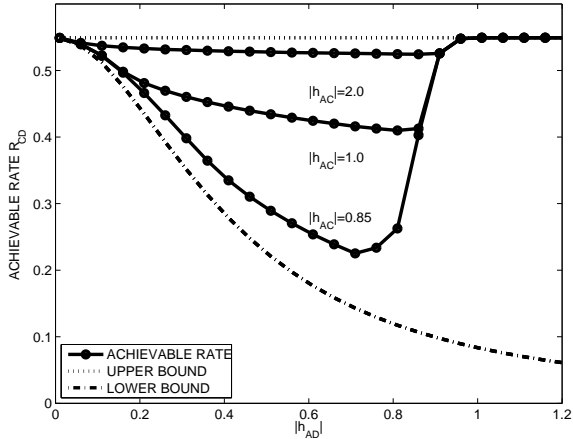}
\caption{Variation of achievable rate with~$|h_{AD}|$ for different
values of~$|h_{AC}|$.} \label{fig:rate_varn_hAD}
\end{center}
\end{figure}
Fig.~\ref{fig:rate_varn_hAD} shows the variation of the achievable
rate~$r_{CD}$ with~$h_{AD}$ for different values of~$h_{AC}$. Notice
the nonmonotonic variation of~$r_{CD}$ with~$h_{AD}$ which can be
explained as follows. First consider~$h_{AC}=$ is small. In this
case, the transmitter cannot reliably decode node~$A$'s signal. If
in addition, $h_{AD}$ is also small, then node~$D$ cannot decode
node~$A$'s signal either. Thus, as $h_{AD}$ increases, the
interference of node~$A$ at node~$D$ increases and the achievable
rate~$r_{CD}$ decreases. Now, as~$h_{AD}$ increases beyond a certain
value, node~$D$ can begin to decode node~$A$'s signal and the
probability of error is captured by Gallager's error exponents. In
this scenario, as $h_{AD}$ increases, the error probability
decreases and thus node~$D$ can cancel out more and more of
interference from node~$A$. Consequently,~$r_{CD}$ increases.
Similar qualitative behavior occurs for other values of~$h_{AC}$.
However, for large~$h_{AC}$, node~$C$ can decode (with some errors)
the signal from node~$A$ and then use a noisy DPC scheme to achieve
higher rates~$r_{CD}$. Notice also that as explained before for
large~$h_{AD}$, the outer bound on the rate is achieved for all
values of~$h_{AC}$.

\begin{figure}[htp]
\begin{center}
\includegraphics[width=3in]{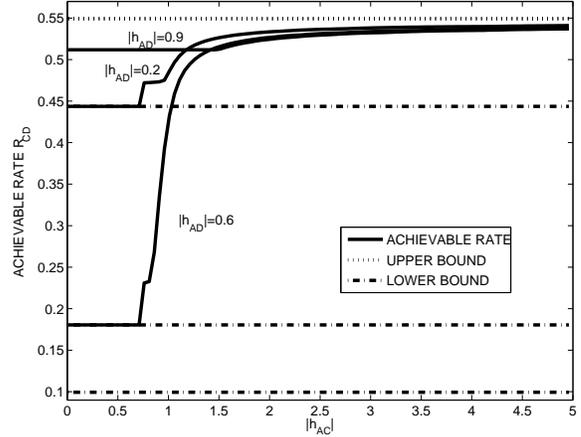}
\caption{Variation of achievable rate with~$|h_{AC}|$ for different
values of~$|h_{AD}|$.} \label{fig:rate_varn_hAC}
\end{center}
\end{figure}
The variation of $r_{CD}$ with~$h_{AC}$ is given in
Fig.~\ref{fig:rate_varn_hAC}. First consider the
case~$|h_{AD}|=0.2$. In this case, node~$D$ cannot decode the signal
of node~$A$ reliably. Now, for small values of $|h_{AC}|$ node~$C$
also cannot decode node~$A$'s signal. Hence, the achievable rate
equals the lower bound,~$R_{CD-lb}$. As $|h_{AC}|$ increases,
node~$C$ can begin to decode node~$A$'s signal and cancel out a part
of the interference using the noisy DPC scheme; hence~$r_{CD}$
begins to increase. Similar behavior is observed for $|h_{AD}|=0.6$.
However, when $|h_{AD}|=0.9$, node~$D$ can decode node~$A$'s signal
with some errors and cancel out part of the interference. Hence, in
this case, even for small values of $|h_{AC}|$ the achievable
rate~$r_{CD}$ is greater than the lower bound. As before~$r_{CD}$
increases with $|h_{AC}|$ since node~$A$ can cancel out an
increasing portion of the interference using the noisy DPC
technique. Note however, that a larger~$h_{AD}$ causes more
interference at node~$D$, which is reflected in the decrease of the
lower bound.  Thus, for a given~$|h_{AC}|$ the achievable rate can
be lower or higher depending on the value of~$|h_{AD}|$.

\bibliographystyle{ieeetr}
\bibliography{DPC_Cog}
\end{document}